\documentclass[preprint,12pt]{elsarticle}

\usepackage{subfig,graphicx}
\usepackage{epstopdf}
\usepackage{subfloat} 
\usepackage{amsmath}  
\usepackage[font=small]{caption}

\usepackage{amssymb}

\journal{Journal of Physics G}

\begin{document}

\begin{frontmatter}



\title{Hadronic observables from a hadronic rescattering model in $Pb+Pb$ collisions at $\sqrt{s_{NN}}=2.76$ TeV}


\author{J. T. Buxton and T. J. Humanic}

\address{Department of Physics, Ohio State University, Columbus, OH, USA}

\begin{abstract}
We employ a simple kinematic model based on the superposition of $p+p$ collisions, relativistic kinematics, and final-state hadronic rescattering to calculate a number of hadronic observables in $\sqrt{s_{NN}}=2.76$ TeV $Pb+Pb$ collisions.  The current model is similar to those used in previous studies, but includes an additional procedure (``squeeze procedure'') which modifies the model pseudorapidity distribution to better represent the experimental data.  In addition, we vary the model hadronization time ($\tau$ = 0.1, 0.2, 0.3 fm/c) to gain a better understanding of our systematic uncertainty.  We find that the simple model describes the overall data reasonably well qualitatively, and in some cases quantitatively.  Furthermore, the model is found to be robust in the sense that using the squeeze procedure and small variations of the hadronization proper time do not significantly affect our results.

\end{abstract}


\end{frontmatter}


\section{Introduction}
The goal of studying relativistic heavy ion collisions is to investigate the fundamental properties of matter at extreme densities \cite{Lisa:2005dd}.  In such collisions, a new type of matter, called the quark-gluon plasma (QGP), is produced in which hadrons dissolve into colored degrees of freedom \cite{Heinz:2013th}.  The QGP behaves as an almost perfect fluid, and this stage of the collision is typically modeled with relativistic viscous hydrodynamics \cite{Heinz:2013th, Muller:2012zq}.  Eventually hadronization occurs, during which the quarks and gluons are once again confined to colorless hadrons, and the system may be described as a hadronic resonance gas \cite{Muller:2012zq}.  

One method of probing the system is to study the numerous hadrons produced in such a collision.  However, the connection between these observables and the early stages of the collision is complicated by random final-state rescatterings.  We attempt to unfold these effects by performing a hadronic rescattering calculation, thereby moving one step closer to the initial state.

The general strategy in building the hadronic rescattering model is to first devise a simple model for hadronization, and then propagate these initial hadrons via rescattering to freeze-out.  As in \cite{Humanic:2010su}, a short hadronization proper time is assumed.  We are then able to extract a parameter set from the model describing the state of the collision before rescattering, and to compare our calculations with experimental data.  The observables studied include spectra (pseudorapidity and transverse momentum distributions), anisotropic flow ($v_{2}$, $v_{2}/n_{q}$, and $v_{3}$ as a function of $p_{T}$), and two-pion femtoscopy (HBT interferometry).  The model includes only hadronic degrees of freedom, and thus should be treated as a limiting case scenario.

The paper is organized as follows:  Section 2 briefly describes the model, Section 3 contains our model results, comparisons to experimental data and discussions for $\sqrt{s_{NN}}=2.76$ TeV $Pb+Pb$ collisions, and Section 4 presents our summary and conclusions.

\section{Description of the Model}
The model consists of five main steps, each to be detailed in the following subsections:  A) generate hadrons in $p+p$ collisions from PYTHIA, B) superpose $p+p$ collisions in the overlap volume of the collision, C) assume a common hadronization proper time ($\tau$), and obtain the position and momentum 4-vectors of the PYTHIA-generated hadrons at hadronization, D) calculate the effects of final-state rescattering among the hadrons, and E) calculate the desired observables.  In this section, we also introduce the new ``squeeze procedure'' (Section 2.6).

\subsection{Generation of the $p+p$ collisions with PYTHIA}
\begin{figure}
  \centering
  \includegraphics[width=100mm]{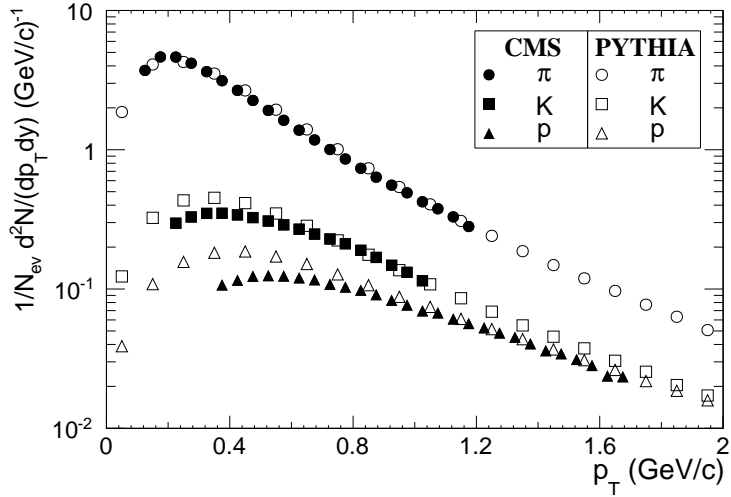}
  \caption{Transverse momentum distributions of identified hadrons (pions, kaon, and protons) at mid-rapidity ($|y|<1$) from PYTHIA compared to CMS data \cite{Chatrchyan:2012qb}.}
  \label{fig0}

\end{figure}

The $p+p$ collisions are modeled with the PYTHIA code \cite{Sjostrand:2006za}, version 6.409, using the internal parton distribution functions ``CTEQ 5L'' (leading order).  The events were generated in ``minimum bias'' mode by setting the low-$p_{T}$ cutoff for the parton-parton collisions to zero, and by excluding elastic and diffractive collisions.  The collisions are run at $\sqrt{s_{NN}}=2.76$ TeV to simulate current LHC data.  The data saved from a PYTHIA event to be input into the next step of the model include the momenta and identities of the ``direct'' (i.e. redundancies removed) hadrons (all charged states) $\pi$, K, p, n, $\Delta, \Lambda, \rho, \omega, \eta$, $\eta$', $\phi$ and K*.  The particles chosen are the most common hadrons produced, and thus should have the greatest effect in our calculations.  Figure \ref{fig0} compares the transverse momentum distribution of identified hadrons from the PYTHIA $p+p$ run (used to generate the $Pb+Pb$ collisions in the present model) to $2.76$ TeV $p+p$ collision data from CMS \cite{Chatrchyan:2012qb}.  As shown, the PYTHIA distribution agrees with the data quite well for pions, but the kaon and proton production are overestimated at low-$p_{T}$.

\subsection{Superposition of $p+p$ events to simulate heavy ion collisions}
An assumption of the model is that an adequate job of describing a heavy-ion collision may be achieved by superposing PYTHIA-generated $p+p$ collisions (at the relevant beam $\sqrt{s_{NN}}$) within the collision geometry.  The collision geometry is the typical smooth ``almond shape'' produced by the overlap volume of the two (Lorentz contracted) disk-like nuclei.  For a collision of impact parameter $b$, the overlap volume (normalized to unity for complete overlap) is described by $f(b)$, such that $f(b=0)=1$ and $f(b=2R)=0$, where $R=1.2A^{1/3}$ and $A$ is the mass number of the nuclei.  The number of $p+p$ collisions to be superposed is then given by $N_{pp}=f(b)A$.  Once the collision geometry is determined, the positions of the superposed $p+p$ pairs are randomly distributed throughout the overlap volume.  The positions are then projected onto the transverse (x-y) plane, requiring all PYTHIA events to occur on the $z_{pp}=$ 0 plane;  the coordinates for a particular $p+p$ pair are defined as $x_{pp}, y_{pp}$ and $z_{pp}=$ 0.  The positions of the hadrons produced in a given $p+p$ collision are defined with respect to the position of the superposed $p+p$ collision center (see Section 2.3).

In calculations performed with a similar model for RHIC collisions \cite{Humanic:2008nt}, better agreement with experimental pseudorapidity distributions ($dN_{ch}/d\eta$) was achieved by imposing a multiplicity cut rejecting the lowest 26\% of $p+p$ collisions.  The justification for this cut was to partially compensate for the fact that primary nucleons from the projectile-target system are not allowed to reinteract in the model.  In the current model, to achieve the same effect, we instead include more $p+p$ collisions for a given geometry, and impose no multiplicity cut.  The scale factor dictating the increase of $N_{pp}$ is determined by matching the model $dN_{ch}/d\eta$ distribution at midrapidity in most central events to that of the experimental data.  For $Pb+Pb$ collisions at $\sqrt{s_{NN}}=2.76$ TeV, this scale factor is 1.35.

Our procedure for generating initial conditions most closely resembles a simplified Glauber Monte Carlo (GMC) approach.  There are, of course, some important differences.  In a typical GMC approach, the nucleons in each of the colliding nuclei are assembled by sampling from a nuclear density distribution.  The two nuclei are aligned according to a random impact parameter (drawn from $\frac{d\sigma}{db} = 2\pi b$), and the collision is then treated as a sequence of independent binary nucleon-nucleon collisions (assuming straight-line trajectories for the nucleons) \cite{Miller:2007ri}.  Both the number of participating nucleons ($N_{part}$) and the number of binary nucleon-nucleon collisions ($N_{coll}$) are then used to determine the initial conditions.  In our model, by randomly distributing the the superposed $p+p$ collisions within the collision geometry, we essentially use a uniform nuclear density distribution.  Since we use only superposed $p+p$ collisions, we track only the binary nucleon-nucleon collisions, and not the individual participants.  Thus, we see that without the rescaling of $N_{pp}$ described above, our initial conditions would be similar to those from a GMC approach in which participants are allowed to undergo only one binary nucleon-nucleon collision.  As such, we scale up the number of $p+p$ collisions to mimic the participants undergoing multiple binary collisions.

\subsection{The space-time geometry picture for hadronization}
The current model only considers hadronic degrees of freedom; however, in the early stages of the collision, partonic degrees of freedom are believed to be more appropriate.  Thus, our calculations must be considered as a limiting case scenario for which a short hadronization proper time is assumed.  However, we do include some effects of the dynamics prior to hadronization by assuming a PYTHIA particle is emitted within a region surrounding the specific $p+p$ collision center, and the particle (``pre-hadron'') travels freely until hadronizing after a proper time $\tau$.  This procedure is explained more clearly in the following.

Consider one superposed $p+p$ collision center (as outlined in Section 2.2) located at a position $(x_{pp},y_{pp},z_{pp}=0)$.  We assume that the PYTHIA particles (from this specific $p+p$ collision) are emitted from a thin uniform disk of radius 1 fm in the $x-y$-plane centered on the collision center (which accounts for the non-vanishing size of the nucleons).  In other words, the emission point $(x_{0},y_{0})$ for a given particle is randomly chosen from a 1 fm disk surrounding the $p+p$ collision center.  Furthermore, we assume that a PYTHIA particle travels freely (as a ``pre-hadron'') until hadronization, which occurs after a time $\tau$ in the particle's rest frame.  We find the model is insensitive to 10\% variations in the radius of the emitting disk.  Similarly, using a Glauber Monte Carlo (GMC) approach, Reference \cite{Alver:2008zza} finds that smearing the produced matter distribution around the GMC interaction points does not significantly influence the observed value of the initial eccentricity (except for extremely small systems).  

The space-time coordinates at hadronization in the lab frame $(x_{h},y_{h},z_{h},t_{h})$ for a particle of momentum $(p_{x},p_{y},p_{z})$, energy $E$, rest mass $m_{0}$, $p+p$ collision center $(x_{pp},y_{pp},z_{pp}=0)$, and transverse disk coordinates $(x_{0},y_{0})$ can be written as

\begin{equation}
\begin{array}{r}
	\vspace{1mm}  
	x_{h}=x_{pp}+x_{0}+\tau\frac{p_{x}}{m_{0}} \\
	\vspace{1mm}  
	y_{h}=y_{pp}+y_{0}+\tau\frac{p_{y}}{m_{0}} \\
	\vspace{1mm}  
	z_{h}=\tau\frac{p_{z}}{m_{0}} \\
	\vspace{1mm}  
	t_{h}=\tau\frac{E}{m_{0}}
\end{array}
\label{eqn1}
\end{equation}

\noindent
Note that the model includes initial expansion in both the transverse and longitudinal directions.  A similar hadronization picture (with an initial point source) has been applied to $e^{+}-e^{-}$ collisions \cite{Csorgo:1990up}.  For the majority of our results, we set $\tau=0.1$ fm/c, as was done in applying a similar model to calculate predictions for RHIC $Au+Au$ collisions \cite{Humanic:2008nt}, Tevatron $p+\overline{p}$ collisions \cite{Humanic:2006ib}, and LHC $Pb+Pb$ collisions \cite{Humanic:2010su}.  Additionally, we study the effects of varying the hadronization proper time by setting $\tau=$ 0.1, 0.2, and 0.3 fm/c.

\subsection{Final-state hadronic rescattering}
The method for calculating the hadronic rescattering is similar to that used in previous studies \cite{Humanic:2010su, Humanic:2008nt, Humanic:2006ib, Humanic:2005ye, PhysRevC.57.866}.  Rescattering is simulated with a semi-classical Monte Carlo calculation which assumes strong binary collisions.  Relativistic kinematics is used throughout.  The hadrons considered in the calculation include pions, kaons, nucleons, and lambdas ($\pi$, K, N, and $\Lambda$), as well as the $\rho, \omega, \eta$, $\eta$', $\phi, \Delta$, and K* resonances.  For simplicity, the calculation is isospin averaged (e.g. no distinction is made among $\pi^{+}, \pi^{0}$ and $\pi^{-}$).

The rescattering simulation proceeds as follows.  Starting from the initial stage ($t=0$ fm/c), the positions of all particles in a given event are allowed to evolve in time in small steps ($\Delta t = 0.5$ fm/c) according to their initial momenta.  At each step, the particle is checked to see if a) it has hadronized, and is therefore able to begin rescattering ($t > t_{h}$, where $t_{h}$ is defined in Eq. \ref{eqn1}), b) it decays, and c) it is sufficiently close to another hadron to scatter.  It is assumed that the two hadrons, $i$ and $j$, scatter when the following criteria are satisfied \cite{PhysRevC.73.054902}

\begin{equation}
\begin{array}{r}
	\vspace{2mm}
	|\Delta \boldsymbol{r}_{c.m.}|_{ij} \leq \sqrt{\frac{\sigma_{ij}}{\pi}} \\
	|\Delta t_{c.m.}|_{ij} \leq t_{0}
\end{array}
\label{eqn2}
\end{equation}

\noindent
where $|\Delta \boldsymbol{r}_{c.m.}|_{ij}$ and $|\Delta t_{c.m.}|_{ij}$ are the separation distance and time difference between the particles in the $i-j$ center of mass frame, $\sigma_{ij}$ is the total scattering cross section for $i$ and $j$, and $t_{0}$ is set to 1 fm/c.  Although a particle may undergo many scatterings, two specific particles are permitted to scatter only once with each other.  Isospin-averaged s-wave and p-wave cross sections for meson scattering are obtained from Prakash et al. \cite{1993PhR...227..321P}, and other cross sections are estimated from fits to hadronic scattering data in the Review of Particle Physics \cite{PDBook}.  Both elastic and inelastic collisions are included.  The rescattering calculation finishes with the freeze-out and decay of all particles.  In practice, the calculation is carried out to 400 fm/c, which allows enough time for all rescatterings to finish.  To test this conclusion, calculations were carried out for longer times, and no changes were found.  Note, after this time is reached, any un-decayed resonances are allowed to decay with their natural lifetimes, and their projected decay positions and times are recorded.

The final-state hadronic rescattering code used in the model resembles a simplified UrQMD model in cascade mode.  The main differences are as follows.  The UrQMD model utilizes a much larger sample of particle species (more than 55 baryon species and 32 meson species \cite{Bass:1998ca}).  In addition, our model is isospin averaged.  Finally, UrQMD includes string fragmentation and excitation \cite{Petersen:2008kb}, which is not explicitly included in our model, resulting in a more string related initial state in the UrQMD model.

\subsection{Calculation of the hadronic observables}
Model runs for $Pb+Pb$ collisions are made to be ``minimum bias'' by having the impact parameters of collisions follow the distribution $\frac{d\sigma}{db} \propto b$, where $0<b<2R$.  The model observables are calculated in a manner typical to experiment by binning the data in centrality through multiplicity cuts.  Note, although we have access to the impact parameter in each event, utilizing multiplicity cuts to determine centrality facilitates our comparison with the experimental data.  In addition, we employ kinematic cuts on pseudorapidity ($\eta$), transverse momentum ($p_{T}$), and pair transverse momentum ($k_{T} = |\vec{p}_{T,a}+\vec{p}_{T,b}|/2$) to duplicate those made in ALICE measurements.  Our analysis focuses on particles emitted near midrapidity ($|\eta| \leq 0.8$).  For the present study, 183,594 minimum bias events were generated from the model for $\sqrt{s_{NN}} = 2.76$ TeV $Pb+Pb$ collisions.

\subsection{The squeeze procedure}
We do not expect the early stages of a nucleus-nucleus collision to behave exactly like the superposition of simple $p+p$ collisions.  For instance, nucleons can scatter multiple times in each nucleus, producing greater ``stopping'', particularly for more central collisions.  In order to approximately account for this effect, and to better represent the experimental pseudorapidity distribution, we use an ad hoc procedure (the ``squeeze procedure'') to modify the pseudorapidity distribution of the particles in our model.  We find that this adjustment does not significantly alter our results for other studied hadronic observables.  The squeeze procedure adjusts the pseudorapidity of a given particle (before input to the rescattering calculation) by implementing the following ad hoc transformation:

\begin{equation}
\begin{array}{l}
	\vspace{2mm}
	\eta' = \eta(1-a \cdot exp[-(\eta-\eta_{0})^{2}/2W^{2}]) \\
	a = a_{0}(b_{max}-b)/b_{max}
\end{array}
\label{eqn3}
\end{equation}

\noindent
where $a_{0} =$ 0.2, $\eta_{0} =$ 4.0, $W =$ 2.0 and $b_{max} =$ 12.10.  The impact parameter ($b$) dependence of the $a$-parameter accounts for the expectation of less stopping for more peripheral collisions (i.e. for larger $b$).

\begin{figure}
  \centering
  \includegraphics[width=135mm]{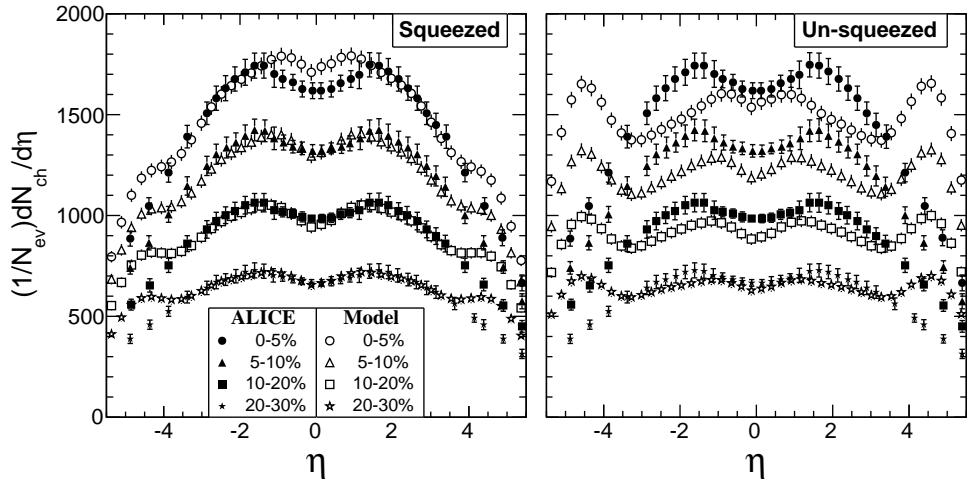}
  \caption{Model charged-hadron pseudorapidity distributions (open markers) for $Pb+Pb$ collisions at $\sqrt{s_{NN}}=2.76$ TeV for the centrality bins 0-5\%, 5-10\%, 10-20\%, and 20-30\%.  Also shown are measurements from ALICE \cite{Abbas:2013bpa} for this energy (closed markers).  Model results in the left panel implement the squeeze procedure, while those in right do not.}
  \label{fig1}
\end{figure}

The effects of the squeeze procedure can be seen in Figure \ref{fig1}, which compares the pseudorapidity distribution of experimental ALICE data \cite{Abbas:2013bpa} to the model calculations with (left panel) and without (right panel) implementation of the squeeze procedure.  Shown in the right panel, without the squeeze procedure the model qualitatively describes the trends of the data near midrapidity, but the model is unable to properly describe the shape of the experimental distributions.  In particular, the un-squeezed model is seen to underestimate the data for $|\eta| \lesssim 3.5$ for all multiplicity bins shown.  Additionally, the relative peak in the experimental distribution around $|\eta| \approx 2$ occurs at smaller absolute rapidity in the un-squeezed model.  Finally, rescattering in the un-squeezed model develops additional ``bumps'' in the pseudorapidity distribution around $|\eta| \approx 4.25$ not present in the data.

As shown in the left panel of Figure \ref{fig1}, once the squeeze procedure is implemented, the model matches the ALICE data very well.  The squeeze procedure shifts all particles toward midrapidity, with the magnitude of the shift determined by the Gaussian term in Eq. \ref{eqn3}.  This Gaussian is centered around $\eta_{0} =$ 4.0, the position for which the $\eta$ shift is maximal.  The squeeze procedure rids our pseudorapidity distribution of the ``bumps'' around $|\eta| \approx 4.25$ and much better approximates the ALICE data.  Figure \ref{fig2} shows how the squeeze procedure modifies the PYTHIA $dN/d\eta$ distribution, prior to the rescattering calculation.

\section{Model Results and Comparisons to ALICE for $Pb+Pb$ at $\sqrt{s_{NN}}=2.76$ TeV}
Results from the model for hadronic observables including spectra, anisotropic flow, and two-boson femtoscopy for $\sqrt{s_{NN}}=2.76$ TeV $Pb+Pb$ collisions are presented separately in the following.  Unless otherwise stated, the squeeze procedure has been implemented and the hadronization proper time taken to be $\tau=0.1$ fm/c.

\subsection{Spectra}
As previously introduced, Figure \ref{fig1} shows model results with the squeeze procedure implemented for the charged-hadron pseudorapidity distributions for various centrality bins.  Also presented are ALICE data for comparison \cite{Abbas:2013bpa}.  The model agreement with the data using the ad hoc transformation in Eq. \ref{eqn3} is seen to be good; however, the model is slightly too high near midrapidity ($|\eta| \lesssim 1$) in the most central collisions, as well as away from midrapidity ($|\eta| \gtrsim 4$) for all studied centralities.  Note, since the model is isospin averaged, the model distributions are multiplied by 2/3 to approximate all charged particles.

\begin{figure}[h!]
\begin{center}
\includegraphics[width=100mm]{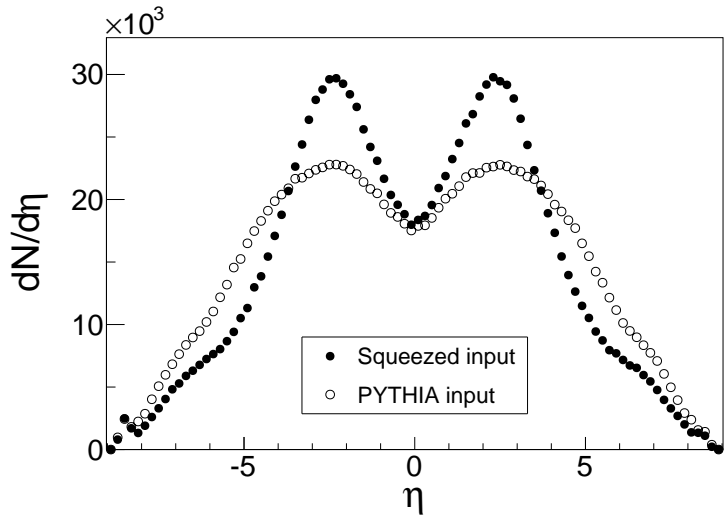} 
\caption{PYTHIA vs. squeezed hadron pseudorapidity distributions before rescattering for $Pb+Pb$ collisions at $\sqrt{s_{NN}}=2.76$ TeV for minimum bias events.}
\label{fig2}
\end{center}
\end{figure}

Figure \ref{fig2} shows the effect of the squeeze transformation on the PYTHIA $dN/d\eta$ distribution.  The ``squeezed distribution'' shown is averaged over all impact parameters.  Even though the transformation in Eq.\ \ref{eqn3} is ad hoc to agree with experiment, it shows qualitatively on average how the $p+p$ pseudorapidity distribution is modified in a $Pb+Pb$ collision.

\begin{figure}[h!]
  \centering
  \subfloat[Large $p_{T}$ range]{
    \label{fig3:subfig:a}
    \includegraphics[width=115mm]{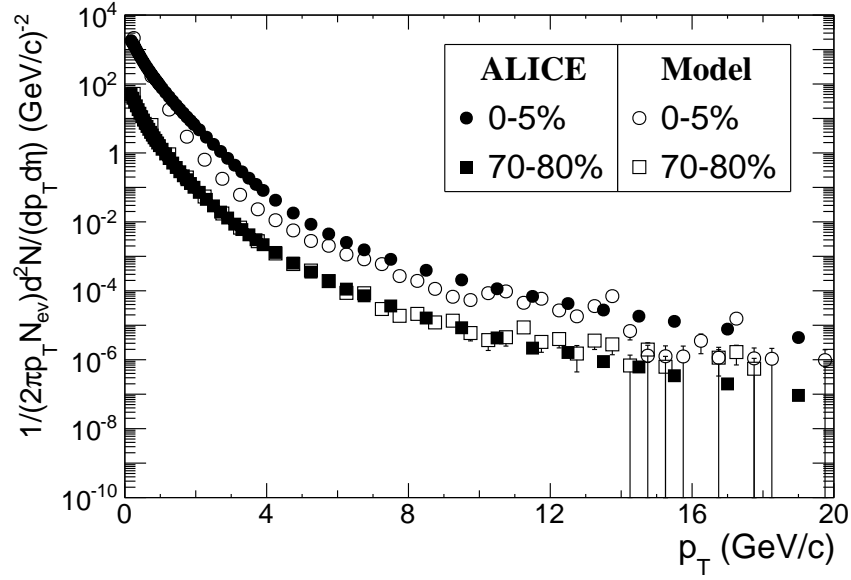}}\\
  \subfloat[Small $p_{T}$ range]{
    \label{fig3:subfig:b}
    \includegraphics[width=115mm]{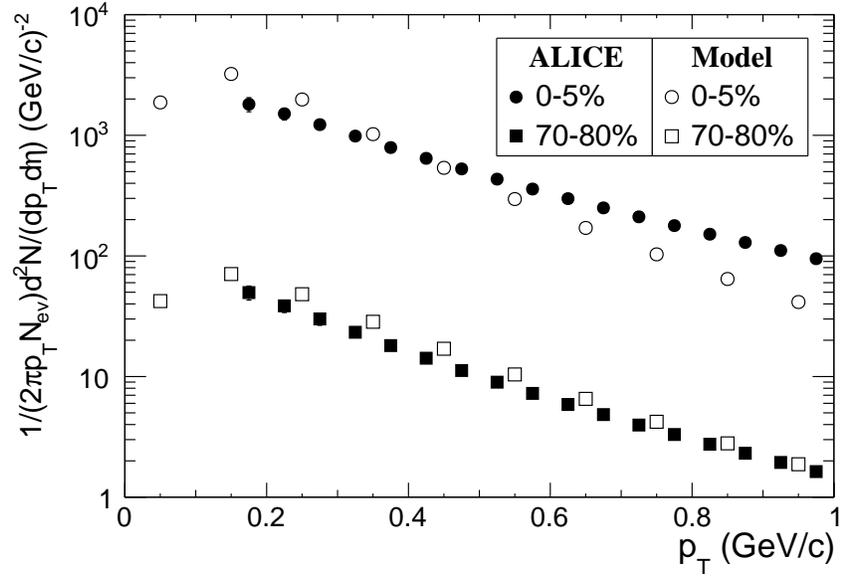}}
  \caption{Model charged-hadron $p_{T}$ distributions compared with ALICE measurements \cite{Abelev:2012hxa} for $|\eta| < 0.8$ and centrality bins 0-5\% and 70-80\%.}
  \label{fig3:subfig}
\end{figure}

Figure \ref{fig3:subfig} shows model comparisons with ALICE data \cite{Abelev:2012hxa} for charged-hadron $p_{T}$ distributions at midrapidity ($|\eta| < 0.8$) for collision centralities 0-5\% and 70-80\%.  Since the model calculations do not distinguish isospin, to approximate all charged hadrons, the model distributions are multiplied by 2/3.  As seen in Figure \ref{fig3:subfig:a}, the model describes the trends of the data, underestimating it in the range $p_{T} \sim$ 1-6 GeV/c for central collisions, and describing the data well throughout the entire $p_{T}$ range for peripheral collisions.  Figure \ref{fig3:subfig:b} shows a magnification of Figure \ref{fig3:subfig:a} in the $p_{T}$ range 0-1 GeV/c where the majority of particles are found.

\begin{figure}[h!]
  \centering
  \includegraphics[width=100mm]{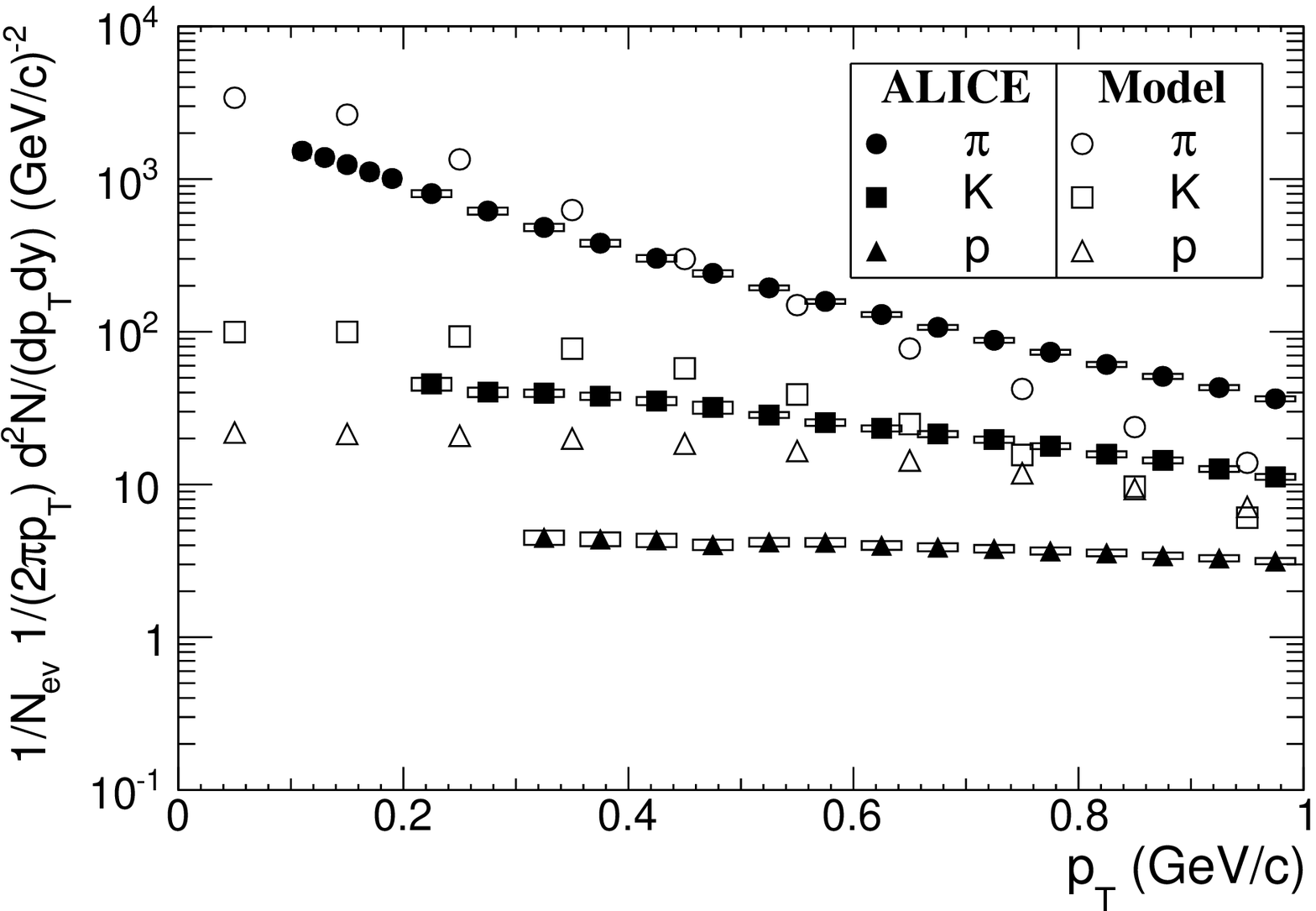}
  \caption{Model transverse momentum distributions of pions, kaons, and protons compared with ALICE measurements \cite{Abelev:2013vea} for $|y|<0.5$ and 0-5\% centrality.}
  \label{fig4}
\end{figure}

Figure \ref{fig4} shows model comparisons with ALICE data \cite{Abelev:2013vea} for the $p_{T}$ distributions of identified pions, kaons, and protons at midrapidity ($|y|<0.5$) in central collisions (0-5\%).  As the model is isospin averaged, to approximate $(h^{+}+h^{-})/2$ the model distributions are multiplied by 1/3 for pions and kaons, and by 1/4 for protons.  The ALICE data has been averaged over positive and negative particles for each species.  As shown, the model reproduces qualitatively the trends of the $p_{T}$ distributions for these identified particles.  However, the pion spectra is much steeper than in the data, and the kaon and proton production are overestimated.

\subsection{Anisotropic flow ($v_{n}$)}
An azimuthal anisotropic flow indicates a collective behavior among emitted particles in a relativistic heavy ion collision, and is observed as an overall pattern which correlates the momenta of final state particles \cite{Muller:2012zq}.  The observed anisotropic flow pattern is typically quantified through a Fourier expansion of the azimuthal distribution of final state hadrons with respect to the collision symmetry planes \cite{Voloshin:1994mz, Poskanzer:1998yz},  
\begin{equation}
	E\frac{d^{3}N}{d^{3}p}=\frac{1}{2\pi}\frac{d^{2}N}{p_{T}dp_{T}dy}\left[ 1+2\sum_{n=1}^{\infty}v_{n}cos[n(\phi-\Psi_{n})]\right]
	\label{eqn4}
\end{equation}

\noindent
where $E$ is the energy of the particle, $p$ the momentum, $p_{T}$ the transverse momentum, $\phi$ the azimuthal angle, $y$ the rapidity, and $\Psi_{n}$ the $n^{th}$ harmonic symmetry plane angle.  The symmetry planes were introduced to account for event-by-event fluctuations of the initial density profile \cite{Abelev:2014pua}.  These symmetry planes are not known experimentally, therefore the anisotropic flow coefficients must be extracted from azimuthal angular correlations between observed particles \cite{Collaboration:2011yba}.   

The study of azimuthally anisotropic flow serves as a sensitive probe of the early evolution of the system, and is typically explained in the language of hydrodynamics as a response of the system to the pressure gradients resulting from initial spatial anisotropies \cite{Heinz:2013th}.  Studying the anisotropy in the final momentum spectrum gains us insights to pressure gradients and density profiles at earlier times.  The evolution of the initial spatial anisotropies to final state momentum anisotropies contains information regarding the equation of state and transport properties controlling the produced matter.

Due to the limited number of particles produced in a given event, one typically averages over events to obtain statistically significant results for the anisotropic flow \cite{Heinz:2013th}.  Fluctuating initial conditions cause the extracted harmonic flow coefficients ($v_{n}$) to fluctuate from event-to-event, even within a fixed and narrow multiplicity bin (or at fixed impact parameter).  In addition, not all azimuthal correlations are collective in origin; these correlations, commonly referred to as ``non-flow'', may result from short-range correlations (like Bose-Einstein effects), resonance decays, Coulomb interactions, and jet correlations \cite{Voloshin:2008dg}.  Therefore, for a correct interpretation of anisotropic flow measurements, one must understand the impact of event-by-event fluctuations and disentangle contributions from non-flow correlations.  Various methods are available for the experimental estimate of the $v_{n}$ coefficients, which depend differently on both flow fluctuations and non-flow correlations \cite{Voloshin:2008dg}.  Utilizing different correlation functions allows one to probe different moments of the $v_{n}$ distributions \cite{Heinz:2013th}.  Two common methods for measuring the anisotropic flow are the event plane method and the cumulant method.

In the event plane method ($v_{n}$\{EP\}) \cite{Poskanzer:1998yz}, the flow is studied by first reconstructing a symmetry plane ($\Psi_{n}$) for the $n^{th}$ harmonic, which, for the second harmonic, is correlated with the (experimentally inaccessible) reaction plane ($\Psi_{RP}$, defined by the beam direction and impact parameter) \cite{Chatrchyan:2012ta}.  The direction of the symmetry planes ($\Psi_{n}$) are determined using the $\phi$-asymmetry generated by the flow itself \cite{Muller:2012zq, Poskanzer:1998yz}.  After a symmetry plane is determined, particle correlations may be formed with respect to it.  Finite multiplicity in each event limits the estimation of the symmetry plane, therefore the flow coefficients ($v_{n}$) must be correctly scaled up by a resolution factor \cite{Poskanzer:1998yz, Voloshin:2008dg}.  Short-range non-flow correlations can be highly suppressed by reconstructing the symmetry plane with particles separated by a large pseudorapidity gap from the particles of interest ($v_{n}$\{EP,$|\Delta\eta|>$2.0\}) \cite{Poskanzer:1998yz, Voloshin:2008dg, Chatrchyan:2012ta, Abelev:2012di}.

The cumulant method ($v_{n}$\{k\}) measures flow by utilizing a cumulant expansion of multiparticle azimuthal correlations \cite{Borghini:2000sa, PhysRevC.64.054901}.  If particles are correlated with the symmetry plane orientation, there should exist correlations between them \cite{Chatrchyan:2012ta}.  Anisotropic flow is a correlation among all particles in an event, whereas non-flow effects arise primarily from few-particle correlations \cite{Voloshin:2002wa}.  Thus, non-flow effects may be suppressed by utilizing a cumulant expansion of multiparticle azimuthal correlations \cite{Borghini:2000sa, PhysRevC.64.054901}.  In practice, measurements utilizing 4-particle correlations ($v_{n}$\{4\}) are shown to suppress non-flow contributions to a negligible level \cite{Collaboration:2011yba, Voloshin:2002wa}.  We utilize the ``direct cumulants'' method outlined in \cite{Bilandzic:2010jr}.  When using two-particle azimuthal correlations, the non-flow effects from short-range correlations can be suppressed by requiring a minimum pseudorapidity separation between correlated particles ($v_{n}$\{k,$|\Delta\eta|>\eta_{sep}$\}) \cite{Voloshin:2008dg}.

\subsubsection{Elliptic flow ($v_{2}$)}
The second anisotropic flow harmonic ($v_{2}$) is known as the elliptic flow, as this component describes a deviation from isotropic emission similar to an ellipse deviating from a circle.  Our analysis differs from experiment in that we have direct access to the reaction plane ($\Psi_{RP}$), defined by the beam direction (z-axis) and impact parameter ($\vec{b}$).  In our model, the coordinates are chosen such that the impact parameter always aligns with the x-axis.  Therefore, the elliptic flow coefficient can also be measured with respect to the reaction plane ($v_{2}$\{RP\}) in the model as

\begin{equation}
\begin{array}{r}
	\vspace{2mm}
	v_{2} = \langle cos(2\phi) \rangle \\
	\phi = tan^{-1}(\frac{p_{y}}{p_{x}})
\end{array}
\label{eqn5}
\end{equation}

where ``$\langle \rangle$'' implies a sum over particles in an event and a sum over events, $p_{x}$ and $p_{y}$ are the $x$ and $y$ components of the particle momentum, and the $x$-axis is in the direction of the impact parameter.  Although this reaction plane method is simpler, we make a more direct comparison to ALICE data by utilizing the event plane and cumulants methods.  Unless otherwise stated, anisotropic flow results are obtained using the event plane method with a minimum pseudorapidity gap of $\eta>$2.0 implemented.

\begin{figure}[h]
  \centering
  \includegraphics[width=140mm]{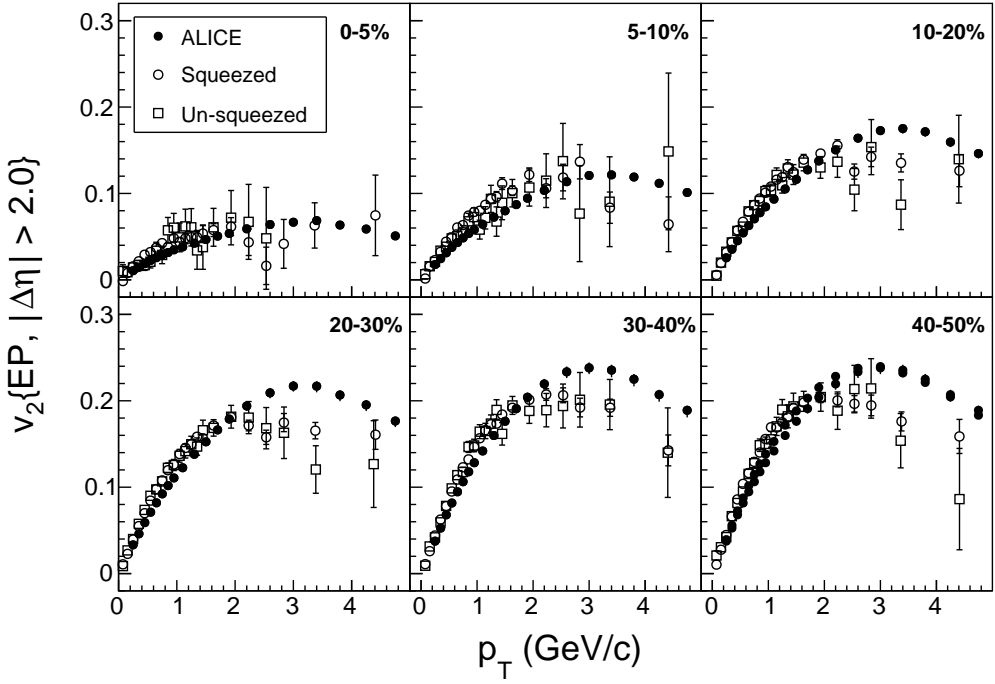} 
  \caption{Model $v_{2}$ vs. $p_{T}$ plots for all hadrons with $|\eta| < 0.8$ for various centrality bins.  Shown are model results with (open circles) and without (open squares) the squeeze procedure implemented, along with ALICE measurements \cite{Abelev:2012di} performed with the event-plane method (closed circles).}
  \label{fig5}
\end{figure}

Figure \ref{fig5} shows the model (with and without implementation of the squeeze procedure) $v_{2}$ vs. $p_{T}$ results for all hadrons at midrapidity for various centrality bins.  For comparison, ALICE data \cite{Abelev:2012di} are shown for $v_{2}$ calculated using the event plane method with a pseudorapidity gap of $|\Delta\eta| >$ 2.0 implemented between particles used in the event plane reconstruction and those of interest.  First, it is interesting to note that the squeeze procedure does not seem to affect the observed $v_{2}$ for $p_{T} \lesssim 1$ GeV/c.  Additionally, the model generates too much elliptic flow at low $p_{T}$, when compared to the ALICE data.  Furthermore, it is remarkable that the model describes the $p_{T}$ behavior of the experiment in which $v_{2}$ increases for low-$p_{T}$, flattens out, and decreases for $p_{T} >$ 3.0 GeV/c.

Large values of elliptic flow are typically considered signatures of the hydrodynamic behavior of the system.  However, the present model, which does not utilize a hydrodynamical description of the system, generates a large amount of flow.  Furthermore, it was found \cite{Humanic:2010su} that the $v_{2}$ signal disappears when the rescattering is turned off in the model, indicating the model flow is due entirely to the hadronic rescattering.  

It is interesting that this model, utilizing a purely hadronic picture, is able to generate such reasonable results.  In our current understanding, we believe the system evolves from an initial hydrodynamic state composed of partons into the hadronic state with possible rescattering and finally freeze-out.  Therefore, it is not unreasonable to assume that the collective effects imprinted in the observables are due to a combination of the hydrodynamic evolution and final-state hadronic rescattering.  Furthermore, the transition from partonic to hadronic degrees of freedom is likely gradual in time, as opposed to a sudden hadronization scenario in a first-order phase transition.  This implies a mixed-phase transition period, during which our simple hadronic rescattering picture could have some degree of validity.  Even at the earliest times, when the degrees of freedom may be purely partonic, hadronic rescattering is possibly able to mimic to some degree the early hydrodynamic evolution of the system. Thus, the purely hadronic rescattering model might be thought of as mimicking a `viscous' hydrodynamic evolution of the system at these early times.  Nonetheless, the current study must be considered as a limiting case picture.

\begin{figure}[h!]
  \centering
  \includegraphics[width=100mm]{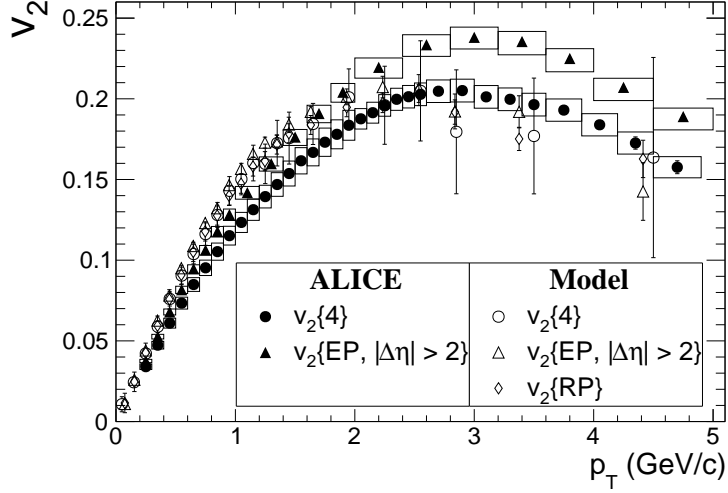}
  \caption{Model $v_{2}$ vs. $p_{T}$ results obtained utilizing various methods for all hadrons with $|\eta| < 0.8$ in the centrality bin 30-40\%.  Shown for comparison are ALICE data \cite{Abelev:2012di} measured with the four-particle cumulant method (closed circles) and with the event plane method (closed triangles).}
  \label{fig6}
\end{figure}

We used a number of different methods to study the elliptic flow.  A comparison of the results from these various methods is shown in Figure \ref{fig6}.  We find, as expected, $v_{2}\{EP\} > v_{2}\{RP\}$ and $v_{2}\{EP\} > v_{2}\{4\}$.  As the four-particle cumulant method reduces non-flow effects, the discrepancy between $v_{2}\{2\}$ and $v_{2}\{4\}$ is typically used to estimate the non-flow contributions in $v_{2}$ measurements.  The difference between $v_{2}\{2\}$ and $v_{2}\{4\}$ in the model is not as large as that in the experimental data, demonstrating a smaller non-flow effect in the model than in experiment.  This is not surprising, as there are no interactions in the model between boson pairs after freeze-out, and Bose-Einstein effects were not introduced here.  Nonetheless, we still have a non-flow contribution from, for example, resonance decays and jet correlations.

\begin{figure}[h!]
  \centering
  \includegraphics[width=100mm]{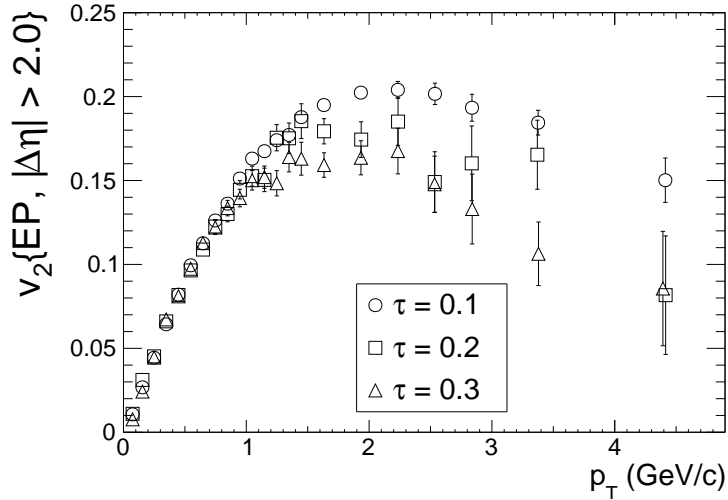}
  \caption{Model $v_{2}$ vs. $p_{T}$ plots for all hadrons with $|\eta| < 0.8$ in the centrality bin 30-50\%.  Shown are model results with three different hadronization proper times assumed:  $\tau=$ 0.1 fm/c (circles), 0.2 fm/c (squares), and 0.3 fm/c (triangles).}
  \label{fig7}
\end{figure}

Figure \ref{fig7} demonstrates the effect on the elliptic flow of varying the model hadronization proper time.  The figure shows $v_{2}$ vs. $p_{T}$ at midrapidity in a centrality window of 30-50\% for the three hadronization proper times $\tau=$ 0.1, 0.2, and 0.3 fm/c.  The variation of the short hadronization time does not much affect the $v_{2}$ signal at low-$p_{T}$ ($\lesssim 1$ GeV/c).  At slightly higher values of $p_{T}$ ($\gtrsim 1$ GeV/c), the curves separate as expected, with the shortest hadronization proper time corresponding to the largest flow signal.  Recall, in the model, a particle (``pre-hadron'') does not scatter until it has hadronized.  Thus, increasing the hadronization proper time increases the average separation between hadrons at the initiation of rescattering, which decreases the amount of rescattering.  As rescattering is responsible for the model $v_{2}$, this leads to a smaller signal.  

\begin{figure}[h]
  \centering
    \includegraphics[width=145mm]{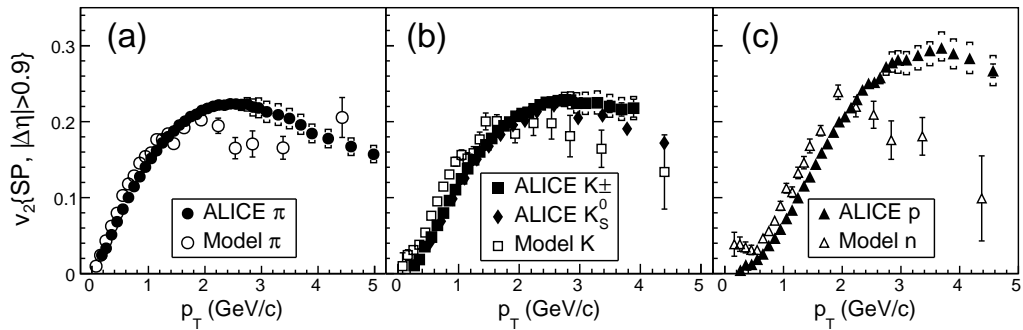}
  \caption{Model $v_{2}$ vs. $p_{T}$ (open markers) for identified (a) pions, (b) kaons, and (c) protons (nucleons in the model) with $|\eta| \leq 0.8$ in the centrality bin 30-40\%.  Shown for comparison are ALICE results \cite{Abelev:2014pua} (closed markers).}
  \label{fig8}

\end{figure}

Figure \ref{fig8} compares the model to ALICE $v_{2}$ vs. $p_{T}$ for identified pions, kaons, and protons (nucleons in the model) at midrapidity for a centrality window of 30-40\%.  For $p_{T} \lesssim 1.5$ GeV/c, the typical mass ordering of the particles is observed, with lower mass corresponding to a higher $v_{2}$ value at a given $p_{T}$.  The model represents the data reasonably well, although at low-$p_{T}$ it slightly overestimates the $v_{2}$ signal for all particle species considered (which is not surprising, when considering Figure \ref{fig5}).  More specifically, the low $p_{T}$ behavior is described well and quantitatively, while the high $p_{T}$ behavior is only described qualitatively.  For all three species of particle, the model $v_{2}$ begins to flatten out at lower $p_{T}$ values than in experiment. The model pion $v_{2}$ matches the data well.  Both the kaon and proton $v_{2}$ at high-$p_{T}$ are underestimated by the model.  The model kaon $v_{2}$ appears more consistent with the experimental $K^{0}_{s}$ data than with the $K^{\pm}$ data.

\begin{figure}[h!]
\begin{center}
\includegraphics[width=90mm]{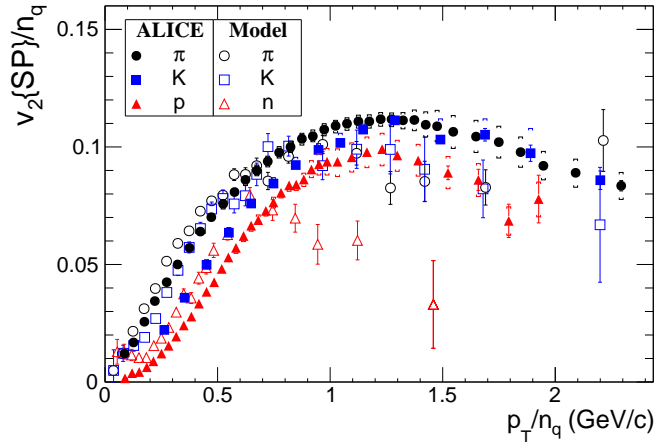} \caption{NCQ scaling of $v_{2}$, i.e. $v_{2}/n_{q}$ vs. $p_{T}/n_{q}$ for identified pions (circles), kaons (squares), and protons (nucleons in the model) (triangles) with $|\eta| \leq 0.8$ in the centrality bin 30-40\%.  Shown for comparison are ALICE results \cite{Abelev:2014pua} (closed markers).}
\label{fig9}
\end{center}
\end{figure}

Figure \ref{fig9} shows the data in Figure \ref{fig8} rescaled by the number of valence quarks in the identified particle ($n_{q}$), as $v_{2}/n_{q}$ vs. $p_{T}/n_{q}$.  This is the so-called ``NCQ scaling'' of $v_{2}$ \cite{Molnar:2003ff}.  If hadronization occurs via coalescence of constituent quarks, there should exist a region in $p_{T}$-space where NCQ scaling approximately holds \cite{Voloshin:2002wa, Molnar:2003ff}.  Such a scaling is typically interpreted as reflecting a collectivity at the quark level, and suggests the system evolves through a phase of deconfined quarks and gluons \cite{Abelev:2014pua}.  The model (without implementation of the squeeze procedure) has been shown to reproduce the apparent scaling observed at RHIC \cite{Humanic:2008nt}.  The current study, with the model scaled to LHC energies, demonstrates a breaking of this apparent scaling, as observed in ALICE data (it appears the apparent scaling observed at RHIC is a coincidence \cite{Muller:2012zq}).  The model is seen to follow the experimental data quantitatively for $p_{T}/n_{q} < 1$GeV/c, and qualitatively at a lower value for $p_{T}/n_{q} > 1$GeV/c.

\subsubsection{Triangular Flow ($v_{3}$)}
\begin{figure}[h!]
\begin{center}
\includegraphics[width=90mm]{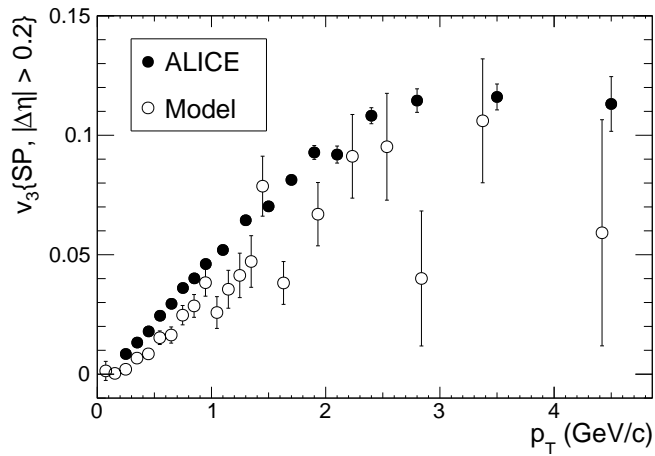} \caption{Model $v_{3}$ vs. $p_{T}$ for all hadrons with $|\eta| \leq 0.8$ in the centrality bin 30-40\%.  Shown for comparison are ALICE data \cite{ALICE:2011ab}.}
\label{fig10}
\end{center}
\end{figure}

The realization of the importance of initial state inhomogeneities and fluctuations has focused much of the recent attention on higher order (and specifically odd) anisotropic flow harmonics.  The third harmonic, triangular flow ($v_{3}$), is driven entirely by fluctuations and lumpiness of the initial density profile.  In general, higher order harmonics are more sensitive to a non-zero viscosity of the expanding system \cite{Muller:2012zq}.  Higher order harmonics are difficult to study due to mode mixing between different order flow harmonics \cite{Heinz:2013th}; however, the triangular flow coefficient is largely free of this effect \cite{ALICE:2011ab}, and therefore serves as an ideal tool for studying both fluctuations and the shear viscosity of the produced system.
Our model produces inhomogeneous and fluctuating initial conditions, so it is natural for us to study the triangular flow.  The model results compared to ALICE data \cite{ALICE:2011ab} for $v_{3}$ are shown in Figure \ref{fig10}.  The triangular flow is analyzed using the scalar products method, which is similar to the event plane method.  The purpose of using this method for $v_{3}$ is to match the method used in \cite{ALICE:2011ab}.  The model reproduces the experimental data well for low-$p_{T}$, and appears to qualitatively describe the flattening $v_{3}$ signal at higher $p_{T}$.  In summary, the past few figures have shown the model to overestimate the elliptic flow ($v_{2}$) and underestimate the triangular flow ($v_{3}$) at low $p_{T}$.  This may be a consequence of our simple initial conditions.

\subsection{Two-pion femtoscopy}
Due to the small size and short lifetime of the system produced in relativistic heavy ion collisions, direct measurements of times and positions is not possible.  Instead, femtoscopy exploits two-particle correlations to help determine spatio-temporal characteristics of such collisions \cite{Lisa:2005dd}.  To perform an experimental femtoscopic measurement, one must measure the two-particle coincident countrate along with the single-particle countrate for reference.  

\begin{figure}
  \centering
  \includegraphics[width=135mm]{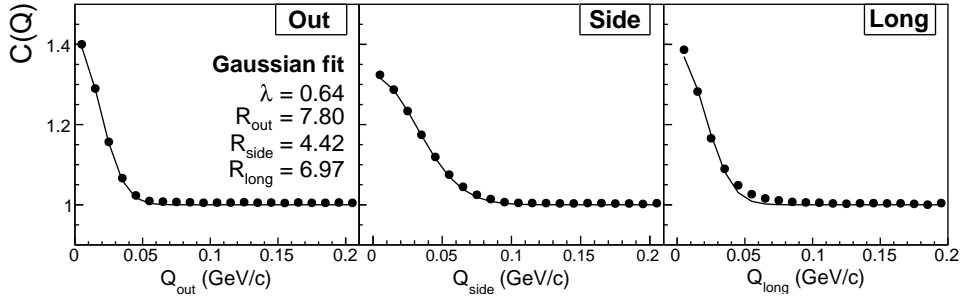}
  \caption{Sample three-dimensional $\pi$-$\pi$ correlation function from the model (points) with Gaussian fit (lines) projected onto the $Q_{out}$, $Q_{side}$, and $Q_{long}$ axes.  The collision centrality is 0-5\% with (single-particle) cuts on the pions $|\eta| \leq 0.8$ and $0.1 < p_{T} < 1.0$ GeV/c, and a pair-cut $0.2<k_{T}<0.3$ GeV/c.  This sample correlation function corresponds to the first $k_{T}$ bin in Figure \ref{fig12:subfig}.  When projecting on one axis, the other two components were required to be less than or equal to 0.03 GeV/c.}
  \label{fig11}
\end{figure}

\begin{figure}
  \centering
  \subfloat[$R_{out}$]{
    \label{fig12:subfig:a}
    \includegraphics[width=65mm]{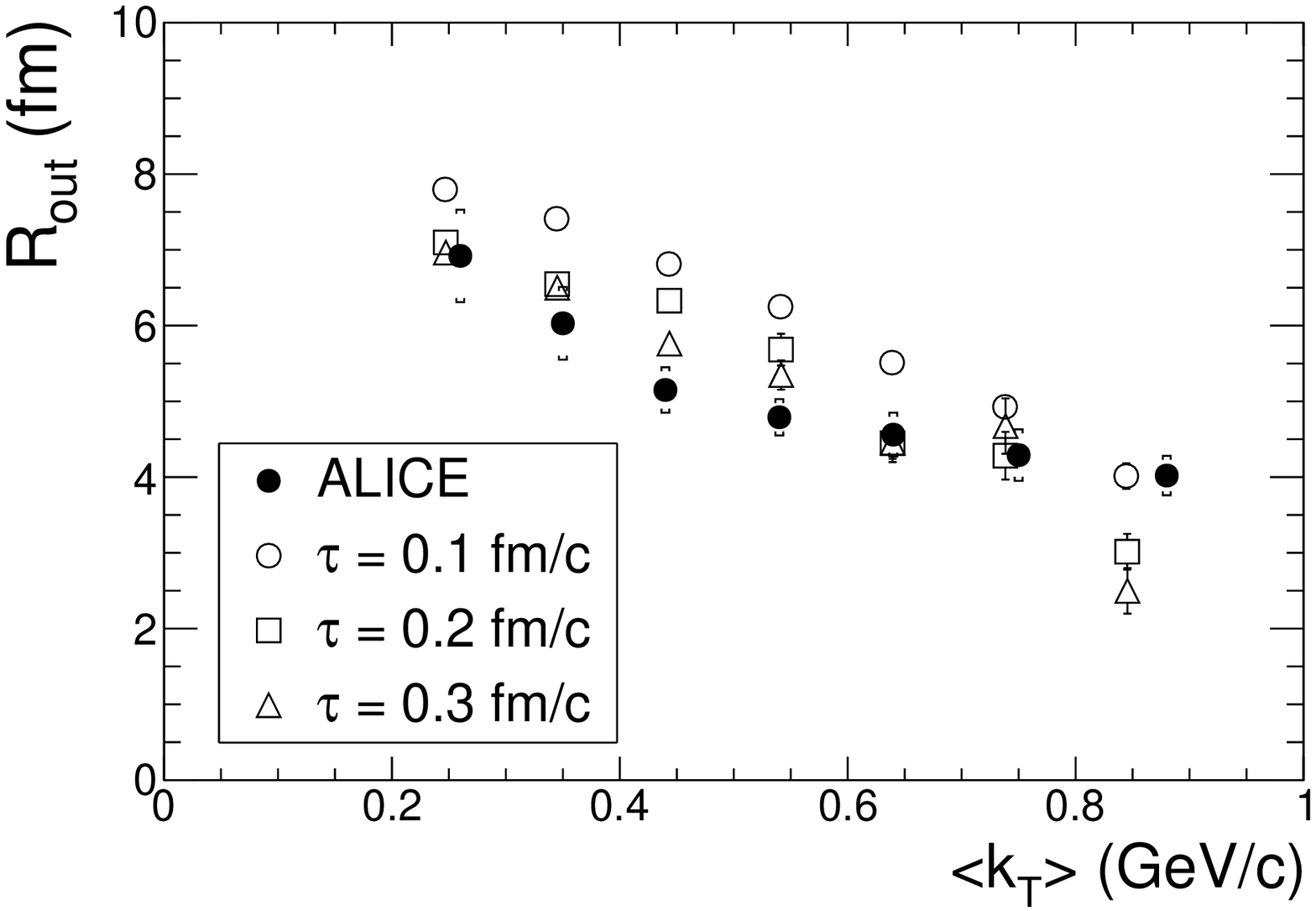}}
  \subfloat[$R_{side}$]{
    \label{fig12:subfig:b}
    \includegraphics[width=65mm]{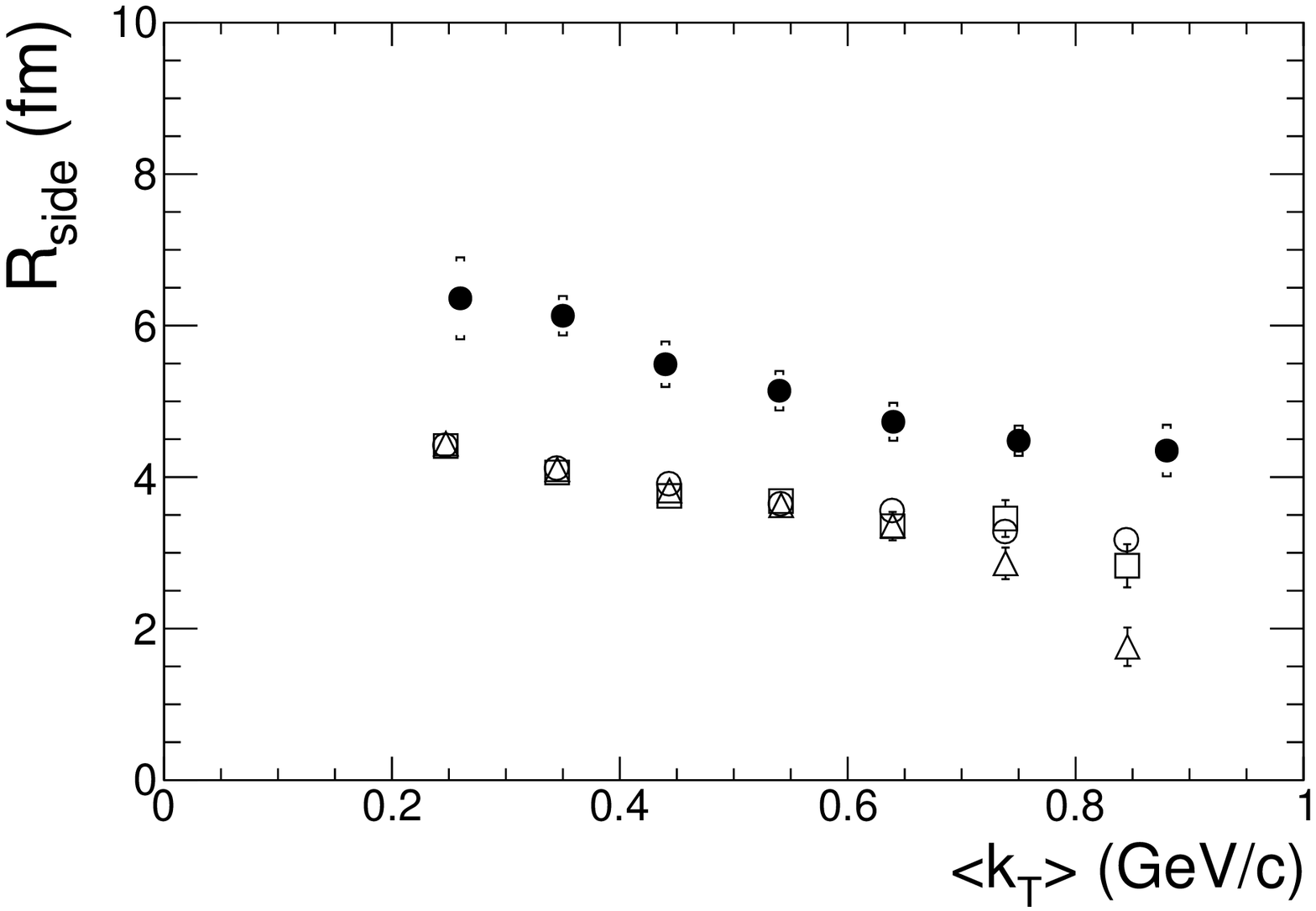}}\\
  \subfloat[$R_{long}$]{
    \label{fig12:subfig:c}
    \includegraphics[width=65mm]{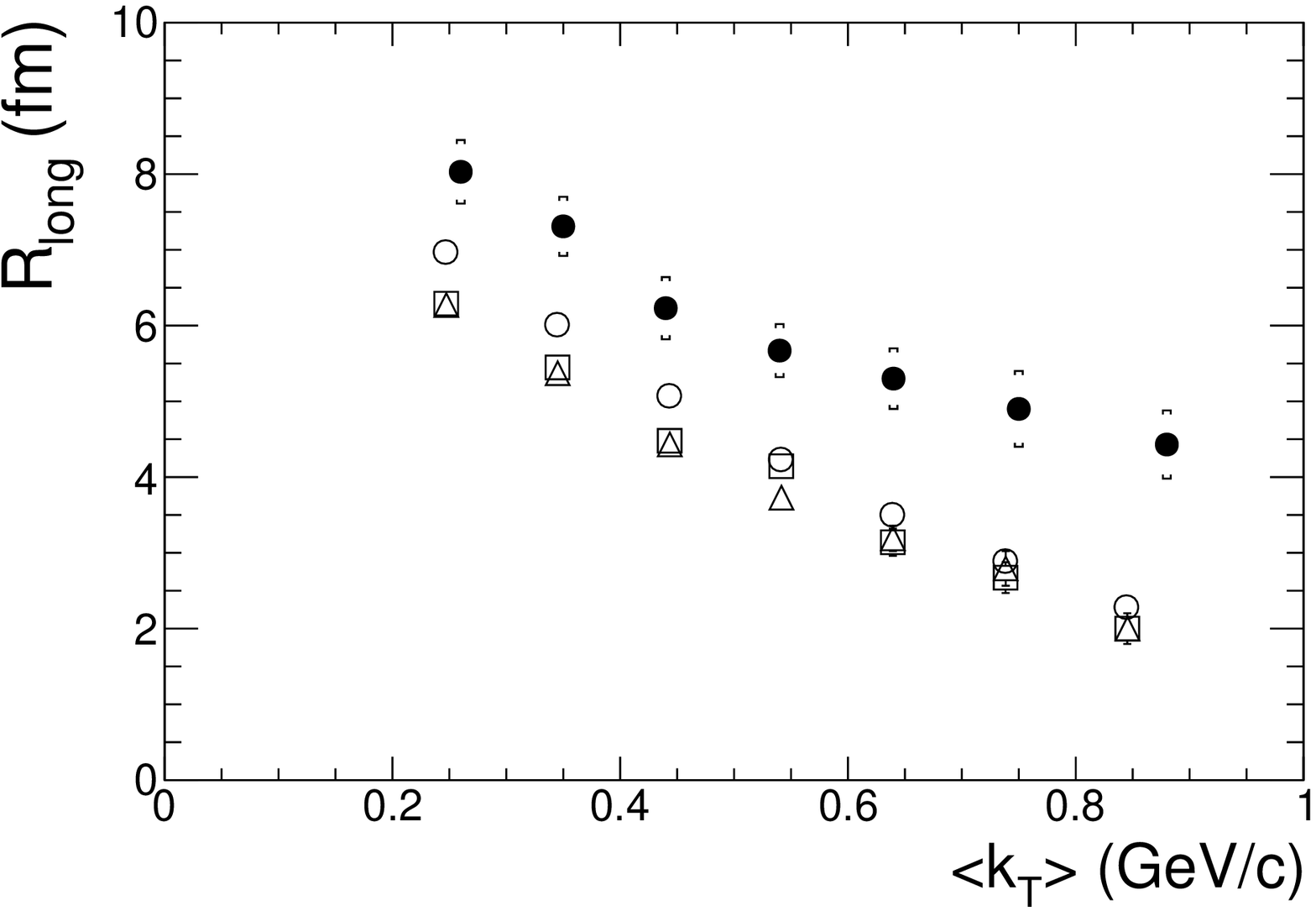}}
  \subfloat[$\lambda$]{
    \label{fig12:subfig:d}
    \includegraphics[width=65mm]{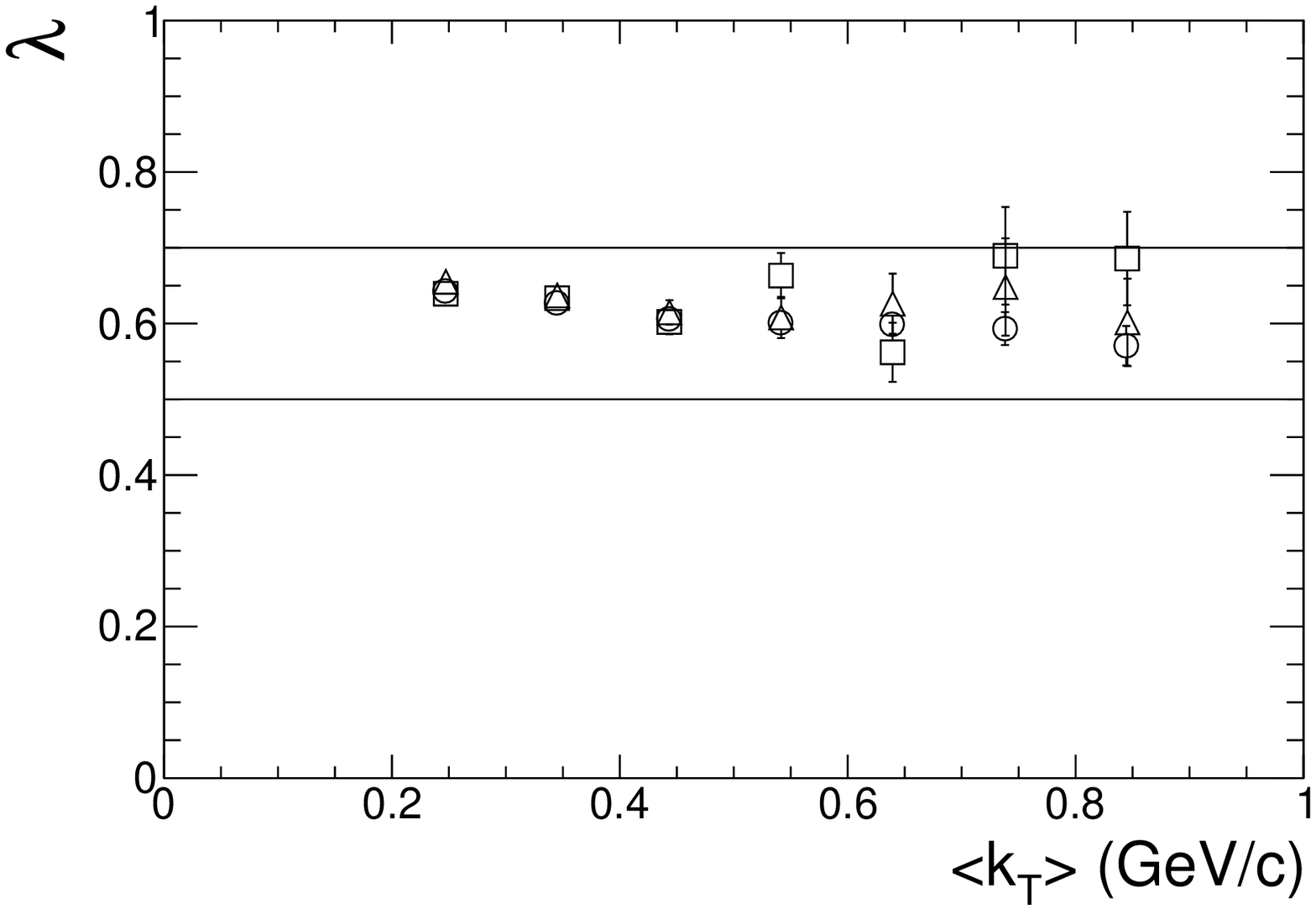}}
  \caption{Model pion source parameters vs. $k_{T}$ for pions at midrapidity ($|\eta| \leq 0.8$) in the top 0-5\% central events.  Shown are model results with three different hadronization proper times assumed:  $\tau=$ 0.1 fm/c (open circles), 0.2 fm/c (open boxes), and 0.3 fm/c (open triangles).  ALICE data \cite{Aamodt:2011mr} (closed circles) are shown for comparison.  In the ALICE data, the error bars represent statistical errors while the error brackets represent the systematic errors.  The horizontal lines in (d) represent the bounds for the $\lambda$-parameters found in \cite{Aamodt:2011mr}.}
  \label{fig12:subfig}
\end{figure}

Figures \ref{fig11} and \ref{fig12:subfig} show results from the model for two-pion femtoscopy for $\sqrt{s_{NN}}=2.76$ TeV $Pb+Pb$ collisions in the 0-5\% centrality range.  To perform the calculation, first the three-dimensional two-boson correlation function is formed from the model data.  Boson statistics are introduced into the model after the rescattering has finished using the standard method of pair-wise symmetrization of the bosons in a plane-wave approximation \cite{PhysRevC.34.191}.  The experimental two-boson correlation function for bosons binned in momenta $\mathbf{k_{1}}$ and $\mathbf{k_{2}}$, $C(\mathbf{k_{1}},\mathbf{k_{2}})$, is constructed from the coincident countrate, $N_{2}(\mathbf{k_{1}},\mathbf{k_{2}})$ and the one-boson contrate, $N_{1}(\mathbf{k})$.  However, it is convenient to express the six-dimensional $C(\mathbf{k_{1}},\mathbf{k_{2}})$ in terms of the three-vector momentum difference, $\mathbf{Q} = |\mathbf{k_{1}}-\mathbf{k_{2}}|$ by summing over momentum difference,

\begin{equation}
	C(\mathbf{Q}) = \sum_{\mathbf{k_{1}},\mathbf{k_{2}}(\mathbf{Q})} \alpha(\mathbf{k_{1}},\mathbf{k_{2}})\frac{N_{2}(\mathbf{k_{1}},\mathbf{k_{2}})}{N_{1}(\mathbf{k_{1}})N_{1}(\mathbf{k_{2}})} = \epsilon(\mathbf{Q}) \frac{A(\mathbf{Q})}{B(\mathbf{Q})}
	\label{eqn6}
\end{equation}

\noindent
where $\alpha(\mathbf{k_{1}},\mathbf{k_{2}})$ is a correction factor for non-HBT effects, $A(\mathbf{Q})$ represents the ``real'' coincident two-boson countrate, $B(\mathbf{Q})$ the ``background'' two-boson countrate composed of products of the one-boson countrates, and $\epsilon(\mathbf{Q})$ is the correction factor for non-HBT effects expressed in terms of $\mathbf{Q}$ \cite{Lisa:2005dd, Heinz:1999rw}.  In practice, $B(\mathbf{Q})$ is the mixed event distribution, which is computed by taking single bosons from separate events \cite{Lisa:2005dd}.  We utilize this procedure to match the experimental analyses.  The pair-wise symmetrization of the pions to account for boson statistics is achieved by weighting the pairs in the coincident countrate by $|\Psi_{ij}|^{2} = b^{2}[1+cos(\Delta k^{\mu} \cdot \Delta r_{\mu})]$, where $\Psi$ is the symmetrized two-pion wave function, $\Delta k^{\mu}$ is the difference in the pair four-momenta ($\Delta k^{\mu} = k_{i}^{\mu}-k_{j}^{\mu}$), and $\Delta r^{\mu}$ is the difference in the pair space-time ($\Delta r^{\mu} = r_{i}^{\mu}-r_{j}^{\mu}$) \cite{Lisa:2005dd, PhysRevC.34.191}.

Since there are no interactions in the model between boson pairs after freeze-out (such as Coulomb or strong interactions), a simple Gaussian function in momentum-difference variables is fitted to Eq.(\ref{eqn6}), allowing the extraction of the boson source parameters which are compared with experiment \cite{Lisa:2005dd},

\begin{equation}
	C(Q_{side},Q_{out},Q_{long}) = G[1+\lambda \cdot exp(-Q^{2}_{side}R^{2}_{side}-Q^{2}_{out}R^{2}_{out}-Q^{2}_{long}R^{2}_{long})]
	\label{eqn7}
\end{equation}
where $\mathbf{Q}$ has been broken up into two transverse and one longitudinal components, the $R$-parameters (radius parameters) are associated with each momentum-difference variable, G is a normalization constant, and $\lambda$ is the usual empirical parameter added to help in the fitting of Eq.(\ref{eqn7}) (for a more complete discussion of the $\lambda$ parameter please see Reference \cite{Lisa:2005dd}).  More specifically, $R_{out}$ points in the direction of the sum of the two boson momenta in the transverse plane, $R_{side}$ points perpendicular to $R_{out}$ in the transverse plane, and $R_{long}$ points in the longitudinal direction along the beam axis.  Note, in the ``ideal HBT case'', $\lambda=1$.  The fit is carried out in the conventional LCMS (longitudinally comoving system) in which the longitudinal boson pair momentum vanishes \cite{Lisa:2005dd}.  Figure \ref{fig11} shows projections of the three-dimensional $\pi$-$\pi$ correlation functions for pairs satisfying $0.2<k_{T}<0.3$ GeV/c (corresponding to the first $k_{T}$ bin in Figure \ref{fig12:subfig}), where $k_{T}$ is the average pair transverse momentum ($k_{T} = |\vec{p}_{T,i}+\vec{p}_{T,j}|/2$).

In Figure \ref{fig12:subfig}, ALICE data \cite{Aamodt:2011mr} are compared to model results assuming three different hadronization proper times: $\tau=$ 0.1, 0.2, and 0.3 fm/c.  In all cases, the model is shown to follow the general trend of the experiment of decreasing radius parameters for increasing $k_{T}$.    We find that the squeeze procedure does not substantially affect any of our results.  The model is seen to qualitatively fit the data well, although it overestimates $R_{out}$ while underestimating both $R_{side}$ and $R_{long}$.  Note that the $\lambda$ parameters mostly all fall within the range $0.5<\lambda<0.7$ found in \cite{Aamodt:2011mr}.  There are two main effects causing $\lambda<1$ in the model.  The first is the presence of long-lived resonances such as $\eta$ and $\eta$' which decay into pions late in the collision, thus suppressing the correlation function.  The second is due to the source deviating from a perfect Gaussian shape.  Note, it was found \cite{Humanic:2010su} that turning off the rescattering in the model, or, similarly setting $\tau >>$ 0.3 fm/c, significantly reduces the HBT radius parameters and mostly eliminates their $k_{T}$ dependences.  Therefore, rescattering also has a strong influence on the HBT parameters in this model.  Additionally, as shown in Figure \ref{fig12:subfig}, the variation of the short hadronization times shown do not have a large effect on our results.

\begin{figure}[h]
  \centering
  \subfloat[$R_{out}$]{
    \label{fig13:subfig:a}
    \includegraphics[width=65mm]{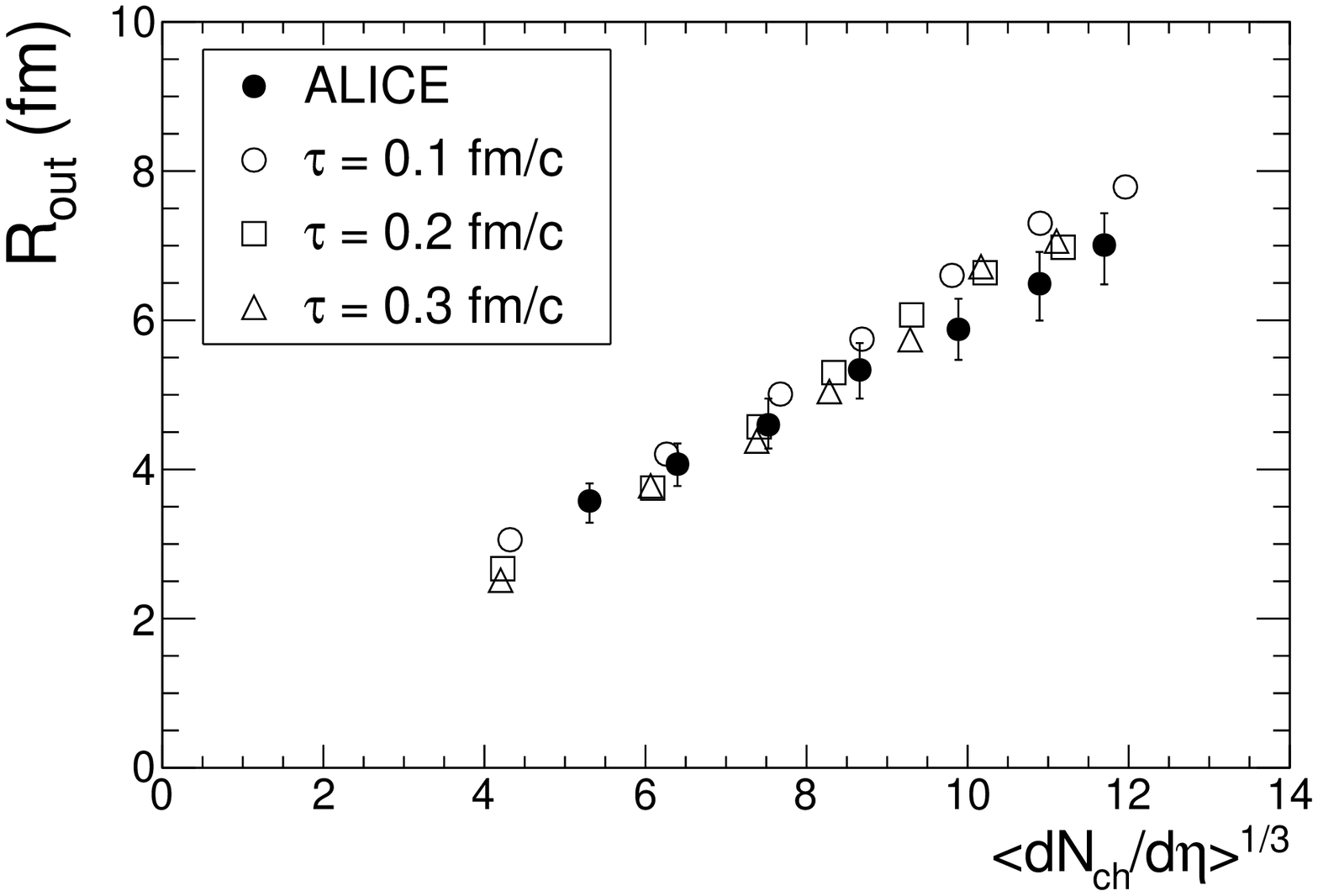}}
  \subfloat[$R_{side}$]{
    \label{fig13:subfig:b}
    \includegraphics[width=65mm]{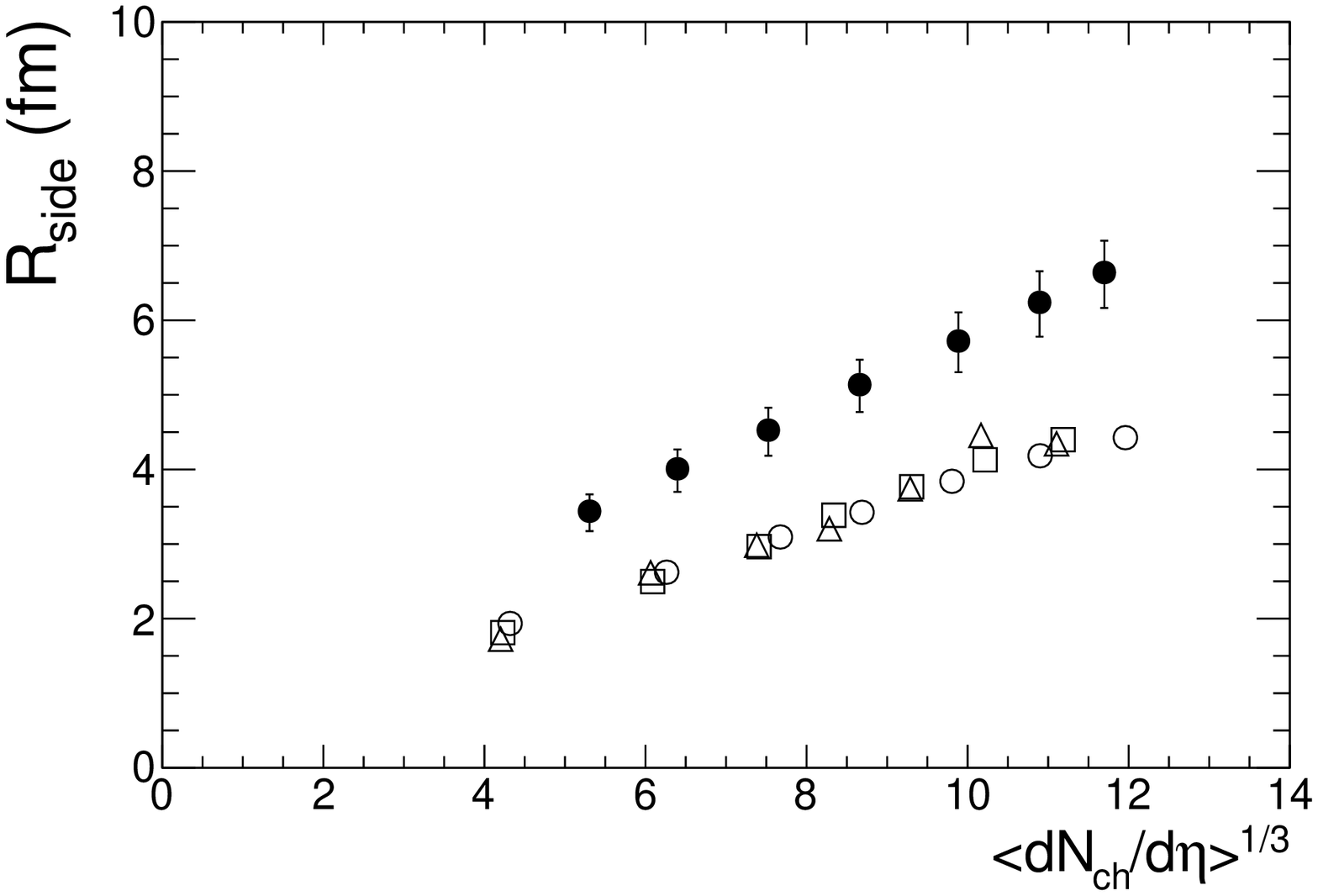}}\\
  \subfloat[$R_{long}$]{
    \label{fig13:subfig:c}
    \includegraphics[width=65mm]{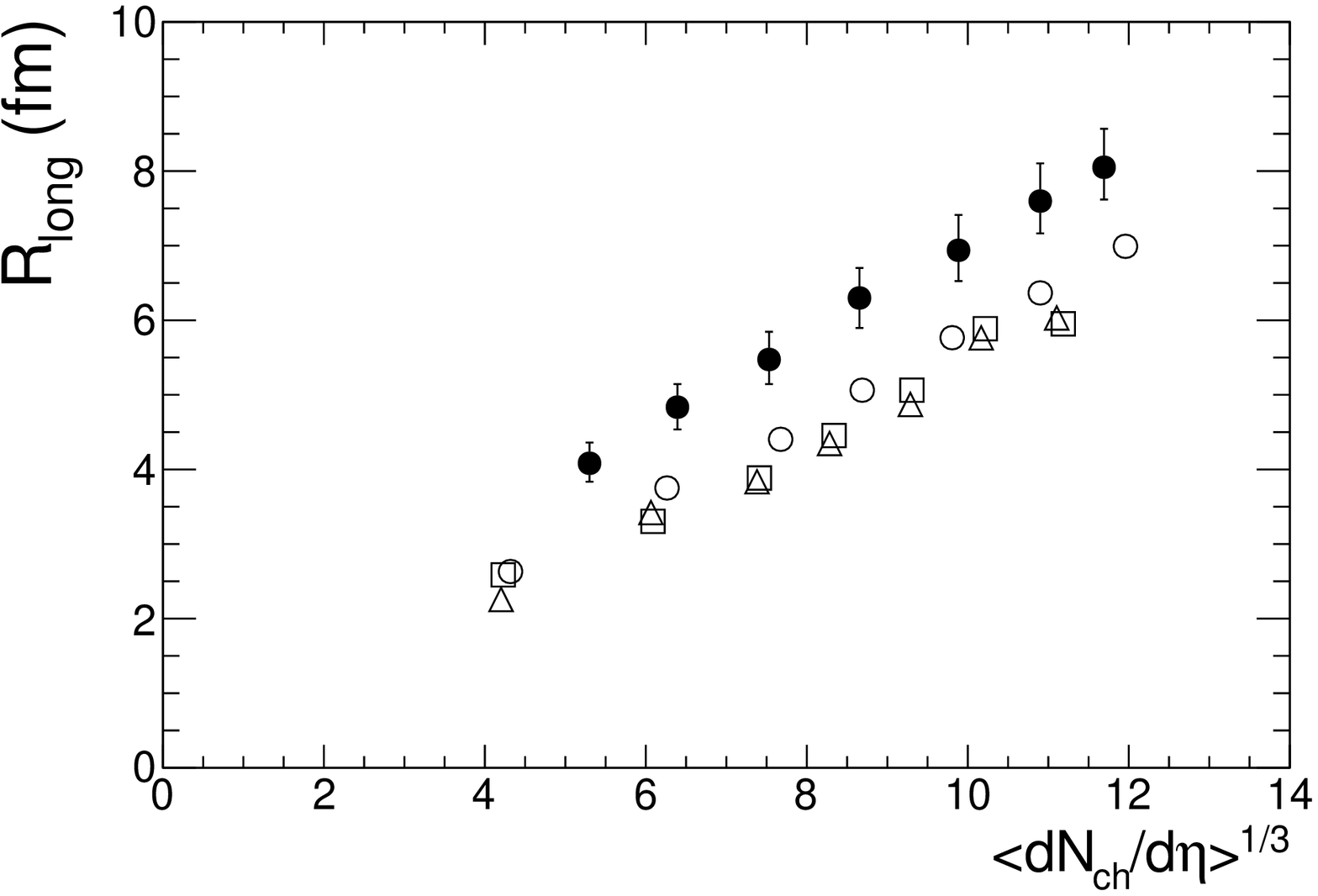}}
  \subfloat[$\lambda$]{
    \label{fig13:subfig:d}
    \includegraphics[width=65mm]{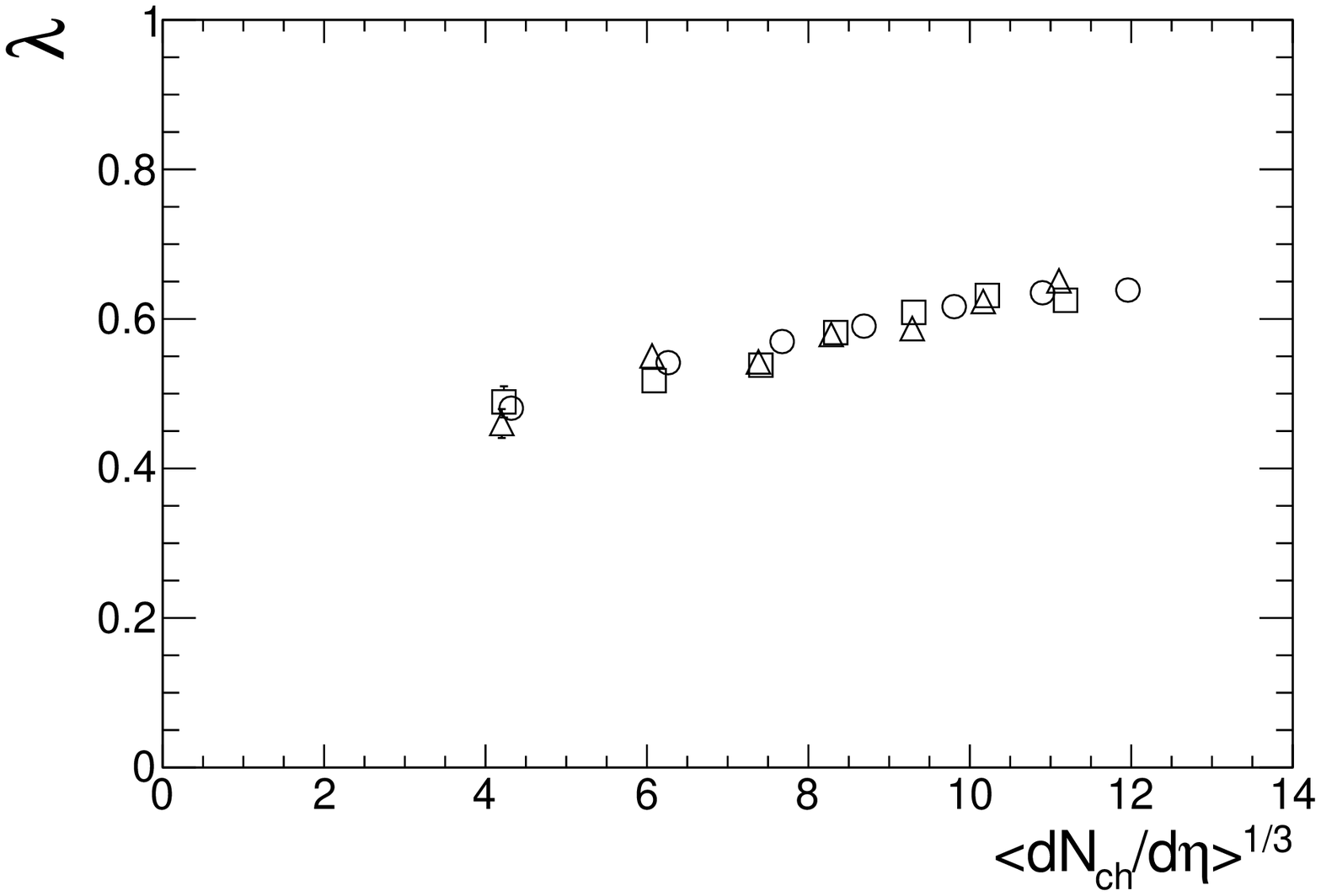}}
  \caption{Model pion source parameters vs. $<dN_{ch}/d\eta>$ for pions satisfying $|\eta| \leq 0.8$ and $0.14<p_{T}<2.0$ GeV/c, and pairs satisfying $0.2 < k_{T} < 0.3$ GeV/c (to match the ALICE data \cite{Graczykowski:2014hoa}).  Shown are model results with three different hadronization proper times assumed:  $\tau=$ 0.1 fm/c (open circles), 0.2 fm/c (open boxes), and 0.3 fm/c (open triangles).  Preliminary ALICE data points \cite{Graczykowski:2014hoa} (closed circles) are shown for comparison.}
  \label{fig13:subfig}
\end{figure}

Figure \ref{fig13:subfig} shows the radii and $\lambda$ parameters for pion pairs satisfying $0.2 < k_{T} < 0.3$ GeV/c plotted as a function of the system size.  To compare with the ALICE data \cite{Graczykowski:2014hoa}, we plot the parameters as functions of $<dN_{ch}/d\eta>$ instead of multiplicity or centrality.  The model data, with decreasing $<dN_{ch}/d\eta>$, correspond to the centralities 0-5\%, 5-10\%, 10-20\%, 20-30\%, 30-40\%, 40-60\%, and 60-80\%, respectively.  The model $<dN_{ch}/d\eta>$ has been approximated using the pseudorapidity distributions (see Figure \ref{fig1}).  As shown in Figure \ref{fig13:subfig}, the model reproduces the expected increase in the radius parameters with increasing multiplicity, i.e. a strong, positive correlation with system size \cite{Lisa:2005dd}.  Similar to Figure \ref{fig12:subfig}, the model describes the data well qualitatively.  Furthermore, the model matches the $R_{out}$ data quantitatively, while it underestimates both $R_{side}$ and $R_{long}$.

\section{Conclusions}
We employ a simple kinematic model based on the superposition of $p+p$ collisions, relativistic geometry, and final-state hadronic rescattering to calculate a number of hadronic observables in $\sqrt{s_{NN}}=2.76$ TeV $Pb+Pb$ collisions.  The model calculations were compared with experimental data from several studies from the LHC.  With the assumption of a short hadronization proper time ($\tau=0.1$ fm/c) in the model, we find that the model describes the trends of the experimental data surprisingly well, when considering its simplicity.  More specifically, we find reasonable agreement with experimental data for spectra (pseudorapidity and transverse momentum distributions), anisotropic flow ($v_{2}$, $v_{2}/n_{q}$, and $v_{3}$), and two-pion femtoscopy.

We find much better agreement with experimental $dN_{ch}/d\eta$ data when implementing our new ``squeeze procedure''.  The squeeze procedure shifts all particles (in a pseudorapidity dependent fashion) toward midrapidity, producing more stopping than in $p+p$ collisions, and aims to partially compensate for multiple interactions of primary nucleons before rescattering.  While greatly enhancing the model agreement with experimental $dN_{ch}/d\eta$, we find that the squeeze procedure does not significantly affect the other studied observables.  However, we focus our study on particles around midrapidity, and a study of the effects of the squeeze procedure away from midrapidity would be interesting.

The main strength of the present model is not a precise agreement for individual observables in a specific kinematic region, but rather its ability to give an overall qualitative description of a range of observables in a wide kinematic region.  Another strength is the simplicity of the model.  The superposed PYTHIA $p+p$ collisions provide all of the information about the initial kinematic state of the hadrons, and the only remaining ``active ingredient'' driving the kinematics underlying the hadronic observables shown is the final-state hadronic rescattering.  Furthermore, the model may be easily scaled to various energies, and has been shown to reasonably reproduce hadronic observables for both $\sqrt{s_{NN}}=2.76$ TeV $Pb+Pb$ and $\sqrt{s_{NN}}=200$ GeV $Au+Au$ collisions.  However, the cost of this simplicity is that we assume that either hadrons or ``hadron-like'' objects can exist in the earliest stage of the heavy-ion collision just after the two nuclei pass through each other.  In other words, the hadronization proper time is assumed short and is insensitive to the environment from which a particle originates.  We do not necessarily believe that hadrons exist at such an early stage, but it is interesting to study this limiting case scenario.  It is interesting to ask why our model gives such reasonable results.  The results presented suggest that our simple hadronic rescattering model is able mimic, to some degree, a 'viscous' hydrodynamic evolution of the system.

\section*{Acknowledgements}
The authors wish to acknowledge financial support from the U.S. National Science Foundation under grant PHY-1307188, and to acknowledge computing support from the Ohio Supercomputing Center.





\section*{References}

\bibliographystyle{jphysg2}
\bibliography{pubbib3}







\end{document}